\documentstyle[epsf]{HC99}

\seceqnum      

\begin{document}

\title[$\beta$ from 2dF]{Estimating $\beta$ from redshift-space distortions in the 2dF galaxy survey}
\author[S.J. Hatton and S. Cole]{Steve Hatton$^{12}$\thanks{hatton@iap.fr} and Shaun Cole$^{1}$\thanks{Shaun.Cole@durham.ac.uk}\\
$^1$Department of Physics, University of Durham, Science
Laboratories, South Rd, Durham DH1 3LE. UK \\
$^2$Institut d'Astrophysique de Paris, 98 bis Boulevard Arago, 
75014 Paris, France.
}

\maketitle

\begin{abstract}
Given the failure of existing models for redshift-space distortions 
to provide a highly accurate measure of the $\beta$-parameter, and the 
ability of forthcoming surveys to obtain data with very low random 
errors, it becomes necessary to develop better models for these 
distortions.  Here we review the failures of the commonly-used 
velocity dispersion models and present an empirical method for 
extracting $\beta$ from the quadrupole statistic that has little 
systematic offset over a wide range of $\beta$ and cosmologies.  This 
empirical model is then applied to an ensemble of mock 2dF southern 
strip surveys, to illustrate the technique and see how accurately we 
can recover the true value of $\beta$.   
We compare this treatment with the error we expect to 
find due only to the finite volume of the survey.  We find that non-linear 
effects reduce the range of scales over which $\beta$ can be fitted, 
and introduce covariances between nearby modes in addition to those 
introduced by the convolution with the survey window function.  
The result is that we are only able to constrain $\beta$ to a $1$-$\sigma$ 
accuracy of $25\%$ ($\beta=0.55 \pm 0.14$ for the cosmological model 
considered).  We explore one possible means of reducing 
this error, that of cluster collapse, and show that accurate 
application of this method can greatly reduce the effect of 
non-linearities, improving the determination of $\beta$.  We conclude 
by demonstrating that, when the contaminating effects of clusters are 
dealt with, this simple analysis of the full 2dF survey yields 
$\beta=0.55 \pm 0.04$.  
For this model this represents a determination of $\beta$ to an
accuracy of $8\%$ and hence an important constraint on the cosmological
density parameter $\Omega_0$.   
\end{abstract}

\begin{keywords}
cosmology: theory -- large-scale structure of Universe --  galaxies: distances and redshifts.
\end{keywords}


\section{Introduction}
\label{sec:intro}

The new generation of galaxy redshift surveys, represented by
the 2dF galaxy redshift survey \cite{Coll95} and the Sloan digital
sky survey (SDSS, \citeNP{gw95}), will be capable of measuring the power 
spectrum and correlation function of galaxy clustering with an 
unprecedented degree of accuracy.  
One way of using this information to constrain the fundamental 
cosmological parameters is through the redshift-space distortions 
of large-scale galaxy clustering.  Since 
the radial coordinate of a galaxy's position is measured by translating 
its redshift to a distance via the Hubble law, the inferred 
distance is prone to contamination by deviations from the 
uniform expansion of the Universe.  This contamination is a 
systematic effect, and produces an anisotropic signal in the 
galaxy correlation function and power spectrum.  The strength 
of this anisotropy depends on the 
mass density of the Universe through the parameter $\beta = \Omega^{0.6}
/b$, where $b$ is the bias factor, relating the amplitude of fluctuations 
in the galaxy density to those in underlying mass \cite{Kaiser87}.

In order to fully exploit the high quality data that will soon 
be available, it is necessary to employ theoretical models that 
are at least as accurate as the data itself.  
In previous work (\citeNP{HC98}, hereafter HC98) we demonstrated 
that two methods often 
used to model the redshift-space distortions in the quasi-linear 
regime were not able to deal robustly with a wide range of cosmological 
parameters.  Their use, when applied to a high-quality dataset, 
could lead to significant bias in determination of the mass density 
of the Universe.  

In this paper we use a large set of \nbody\ simulations, described 
in \citeNP{CHWF} (hereafter CHWF), spanning a range of cosmologies 
and biased in a variety of ways, to show that none of the velocity dispersion 
models commonly considered is capable of estimating $\beta$ without 
substantial bias.  In section~2
we present an empirical model, based on results from these simulations,  
that provides an accurate way of extracting $\beta$ from redshift-space 
distortion information over the full range of cosmologies used.  

In section~\ref{sec:mock}, we address the question of how well the new
galaxy redshift surveys will measure the distortion parameter $\beta$.  
We use our empirical model from section~\ref{sec:model}  to fit 
the results of analysing mock 2dF redshift surveys, whose 
construction is described in CHWF.  Repeating this procedure 
for several independent simulations of the 
same cosmology we estimate the scatter in estimated $\beta$ values 
and hence the accuracy to which we will be able to measure $\beta$.  
We compare this accuracy to that which 
we would expect to find if the density field were 
described by a Gaussian random field, and the 
only source of correlation between modes came from the survey 
window function.  We show that this approximation results 
in a substantially smaller value for the error expected on 
$\beta$.  We attribute this discrepancy to the presence of 
extremely non-linear regions in the mock catalogues which 
cause coupling between modes.  

In section~\ref{sec:clus} we employ the technique of cluster 
collapse to effectively dampen the non-linearities present in 
non-linear regions.  This method greatly 
reduces the error on $\beta$, by extending the range over which 
our models are valid, and by removing some of the mode coupling.
  
In section~\ref{sec:mag} we estimate the tightest possible constraint 
that we will be able to obtain from application of these methods 
to the 2dF survey, by using the the full, magnitude limited sample, 
rather than a volume limited subsample.    
We summarise and conclude in section~\ref{sec:conc}.  

\section{Modelling the redshift-space distortions}
\label{sec:model}
In HC98 we considered two extensions to the linear 
theory of redshift-space distortions: the Zel'dovich 
approximation, which was found to break down in the 
case of biased galaxy distributions, and the exponential 
model of velocity dispersions, which seemed to work well 
only in the limit that the velocity dispersions were high.    
The failure of these existing models to cope with a full 
range of cosmologies and biases is what motivates us 
to find a more accurate and robust model.  

Here we use a set of \nbody simulations spanning a broad range of 
parameter space, with a number of different biasing schemes, 
to look for an empirical function that will accurately model 
the strength of the quadrupole distortion of the galaxy power spectrum.  
We assume that an estimator which can accurately and robustly 
measure $\beta$ from this broad range of simulations should be 
able to do the same when applied to a real galaxy sample.  
These simulations are exactly those described extensively in CHWF, 
but we will summarise their details here.  
We include both open ($\Lambda_0 = 0$) and flat ($\Omega_0 + 
\Lambda_0 = 1$) cosmologies, with power spectra given by 
variants of the CDM spectrum, parameterised in \citeNP{BBKS}.  
The amplitudes 
are set according to two methods, either so as to reproduce 
the local abundance of rich galaxy clusters \cite{WEF93,ECF96}, 
or to match the large-scale \COBE\ observations \cite{BW95}.  
We introduce bias to the simulations by selecting particles as 
galaxies depending on their local density, so as to produce 
a clustering amplitude matching that seen in the APM 
survey \cite{MES96}.  This bias is achieved using a range of different 
prescriptions, as described in CHWF, section~3.3.  These 
prescriptions include high-peaks, power-law and sharp-threshold 
bias models.  The simulations
are run with $192^3$ particles,  on a grid of length $345.6 \hmpc$.  
We select $128^3$ of these particles as galaxies using the biasing 
algorithm.  

\begin{figure}
\centerline{\epsfxsize= 8 truecm \epsfbox{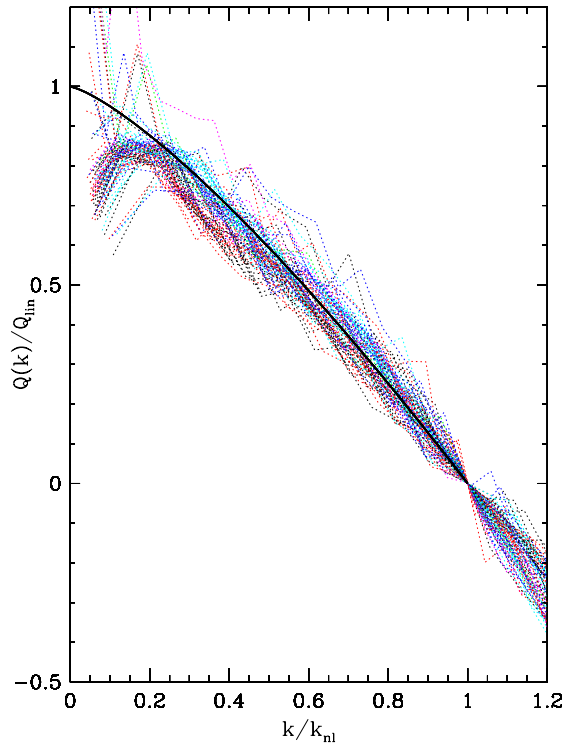}}
\caption{The scaled quadrupole-to-monopole ratio for all the 
simulations and bias models described in CHWF.  The 
$y$-axis is scaled to the expected, linear theory value of 
$Q$ (equation~\protect\ref{eqn:qlin}), the $x$-axis is scaled 
to the zero-crossing scale of the quadrupole, $\knl$.  
The thick black line is our empirical fit, equation~\protect\ref{eqn:fit}.}
\label{fig:scale}
\end{figure}

We expect the redshift-space distortions to tend asymptotically 
towards linear behaviour on the largest scales.  The statistic 
we use to parametrise the distortions is the 
quadrupole-to-monopole ratio, $Q(k) \equiv P_2(k)/P_0(k)$, 
where $P_0(k)$ and $P_2(k)$ represent the monopole and quadrupole
terms of the 3-dimensional redshift-space galaxy power spectrum $P({\bf k})$. 
In the linear regime, a positive value of $Q(k)$ is produced by
coherent infall onto overdense regions and outflow from underdense regions.
The distortion, computed first by Kaiser \shortcite{Kaiser87}, 
depends only on $\beta$  and the resulting expression for the 
quadrupole-to-monopole ratio is
\[
\label{eqn:qlin}
\Qlin = \frac{4\beta/3 + 4\beta^2/7}{ 1 + 2\beta/3 + \beta^2/5 }
\]
(\eg \citeNP{CFW94}).  
Most physical models of local galaxy bias result in a bias 
function, $b$, that, despite being scale dependent on small scales, 
tends to a constant value on large scales \cite{Coles93,MPH98}.  
Thus, $\beta$ itself is expected to be constant in this regime, 
and, as a starting point, any model of $Q(k)$ should have this 
asymptotic behaviour on large scales.  

On smaller scales the quadrupole-to-monopole ratio is suppressed from 
its linear value.  Eventually, $Q(k)$ becomes negative, where the
random motions inside highly non-linear overdense regions produce 
structures elongated along the redshift direction (fingers of God).
Most of the information that can be used to constrain $\beta$ 
comes from the quasi-linear regime, where $Q(k)$ is suppressed but 
still positive.  We therefore seek a model for $Q(k)$
over just this range of scales.  

We measure $Q(k)$ from the simulations using the technique 
of fast Fourier transforms, presented in HC98.  
We first simulate redshift-space anisotropy by picking one  
of the coordinate axes of the simulation as a redshift direction, 
and perturbing all the particle positions by their velocities in 
this direction.  As noted in HC98, using a single line 
of sight for the simulation 
ensures that the distant observer approximation holds, 
by effectively placing the simulation at an infinite distance.   
The galaxy density field, $\delta(\r)$, is Fourier transformed 
to obtain $\delta(\k)$, and $\hat{P}(\k) \equiv \< \delta(\k)^2 \>$
is taken as the estimate of the power spectrum at each grid point 
in \kspace.  This three-dimensional, anisotropic power spectrum is 
then decomposed into spherical harmonics to obtain the monopole 
(or spherically averaged power), $P_0(k)$, and quadrupole, $P_2(k)$.  

Fig.~\ref{fig:scale} shows the quadrupole-to-monopole ratio, 
$Q(k)\equiv P_2(k)/P_0(k)$, for all the galaxy distributions 
resulting from 
each cosmological model and local biasing prescription investigated 
in CHWF.  We have scaled the curves such that the $x$-axis is 
measured in units of the wavenumber, $\knl$, where the quadrupole 
crosses zero, and the $y$-axis is scaled by the expected, linear 
theory value given by equation~\ref{eqn:qlin}.  

At low $k$, there is substantial scatter between the curves, 
and a turn down from the linear value in the first two bins.  
This feature was explained in CHWF as resulting from a random 
down-turn in power on these scales in the simulations we used: 
the majority of the simulations are based on the same initial 
random phases, so this down-turn is repeated in all the simulations.  

Apart from this trend, it can clearly be seen that the curves 
generally have a common locus once they have been scaled 
in this way.  It is thus reasonable to seek a model for $Q(k)$
of the form $Q(k)=\Qlin f(k/\knl)$.  A good empirical fit of this form
is provided by 
\[
\label{eqn:fit}
f(x) = 1-x^{1.22}, 
\]
which is shown by the heavy solid line in Fig.~\ref{fig:scale}. 

We have thus found a model with two free parameters, 
$\beta$ and $\knl$, which accurately describes the shape 
of the quadrupole-to-monopole ratio, $Q(k)$, for the wide
range of models considered in CHWF.  

This is rather similar to the approach used by \citeN{FN96} 
who fit an empirical curve to their analytic Zel'dovich 
approximations for the quadrupole using the same scalings.  
We stress that our approach should be far more general since 
it is based on the fully non-linear data provided by \nbody 
simulations, covers a much broader range of cosmologies, and 
includes the effects of bias.

\subsection{Definition of bias}

\begin{figure}
\centering
\centerline{\epsfxsize= 8 truecm \epsfbox{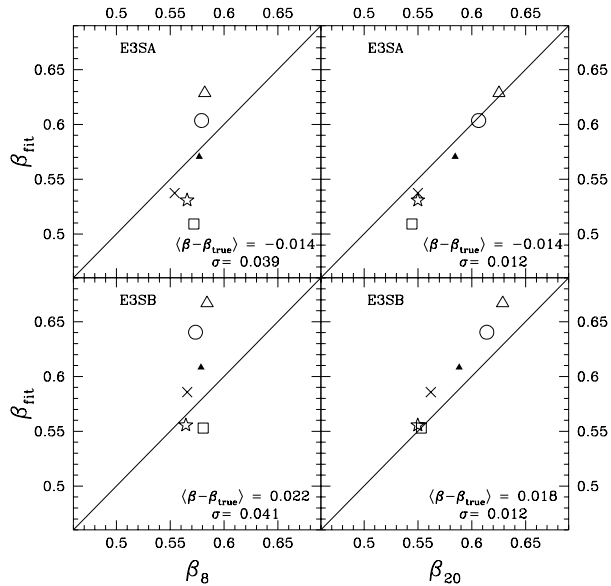}}
\caption{Scatter plots for two realizations of the $\Omega_0 = 1$, 
$\tau$CDM, simulations (named E3SA and E3SB in CHWF), biased with 
each of the six biasing prescriptions described in CHWF.  The point 
types reflect different bias models: model~1 (stars), 2 (crosses), 
3 (circles), 4 (open squares), 5 (solid triangles), 6 (open triangles).
On the left we show comparisons of the fit $\beta$ with the input 
$\beta = \Omega^{0.6}/b$, where the bias factor, $b$, is 
defined by the ratio of the 
fluctuations in spheres of radius $8\hmpc$ in the galaxy 
and matter distributions.  On the right, the bias factor 
is instead defined on a larger scale, as the ratio of fluctuations in 
spheres of radius $20\hmpc$.}
\label{fig:scat2}
\end{figure}

The conventional method for specifying the bias factor, $b$, 
is by the ratio of rms fluctuations 
in spheres of radius $8\hmpc$ in the galaxy density to those 
in the mass density.  
In the approximation that both power spectra have the same shape, 
or at least the same shape on scales that contribute to the 
fluctuations in $8\hmpc$ spheres, this is equivalent to a boost 
in the galaxy power spectrum normalization by a factor $b^2$ 
relative to the mass spectrum.  Most physical prescriptions for 
galaxy formation, in contrast, result in a bias that is to some 
extent scale dependent \cite{MPH98,Benson99,Blanton99}.  For 
example, the large-scale galaxy 
distribution may have a constant bias, but this may be reduced in 
clusters where the number density is sufficiently high that a 
significant amount of merging has occurred.

In Fig.~\ref{fig:scat2} we show the result of fitting our empirical
model to the two simulations 
for which the widest range of biasing schemes is available.  
These $\tau$CDM simulations (named E3SA and E3SB in CHWF)
have been biased using all six prescriptions described in CHWF.  
The left hand panels show the scatter of $\beta_{\rm fit}$ versus 
$\beta_8 = \Omega_0^{0.6}/ b_8$.  The diagonal line represents the 
ideal, one-to-one correspondence between the estimated and 
true values of $\beta$.  The labels at the bottom of each 
panel give the average difference between the fitted and 
true values, and also the scatter around the mean relation.  
As can be seen, for any one simulation, most of the 
biasing prescriptions result in the same values of $\beta_8$; not 
really surprising, since most of them were fixed to reproduce 
clustering on these scales.  Different prescriptions for biasing 
the galaxy catalogues, however, result in systematically different 
best-fitting values of $\beta$, despite the constraint that $\beta_8$ 
is the same.  This produces significant scatter in the values of 
$\beta_{\rm fit}$ for these different biases.  

In contrast, the scatter is much reduced in the right
hand panels in which we plot $\beta_{\rm fit}$ 
against $\beta_{20}$, where $\beta_{20}$ is obtained 
using the bias factor, $b_{20}$, defined by the ratio 
of galaxy to mass fluctuations in spheres of radius $20\hmpc$.  
Ideally we would like to define the bias parameter
on very large scales which would truly reflect the asymptotic 
value of the bias, but in practice the finite volume of the 
simulations makes estimates of the bias on such scales rather noisy.
CHWF presented a method for calculating analytically the asymptotic 
value of the bias, but this method can only be applied to the
subset of bias models that are based on the initial linear 
density field rather than the evolved non-linear density field.
We expect $b_{20}$ to be a good approximation to the asymptotic 
large scale bias in all cases.

Use of $\beta_{20}$ rather than $\beta_8$ has a dramatic 
effect on the scatter of the different models, pulling them 
much closer to the $\beta = \beta_{\rm fit}$ line.  The 
scatter is reduced by a factor of four in these models.  
This tighter correlation indicates that the quadrupole, 
despite being measured over a range of mostly quasi-linear 
wavenumbers, constrains quite directly the large-scale
linear bias factor.

\subsection{Comparison with existing models}
We now compare our two-parameter model for $Q(k)$ with the
analytic velocity dispersion models which are commonly used
in the analysis of redshift-space distortions.  In fact, 
three models of velocity 
dispersion have generally been considered in the literature: 
\begin{enumerate}
\item {\bf Gaussian}.  Particle velocities are drawn from the 
distribution $\prob(v) \propto \exp(- v^2 / 2\sigma_v^2)$.  
In this case the convolution in real space results in a 
multiplication of $\delta(\k)$ by the factor 
$\exp( - k^2 \mu^2 \sigma_v^2 /2)$, where $\mu$ is the 
cosine of the angle between the wave-vector, ${\bf k}$, 
and the line of sight.  

\item {\bf Exponential}.  The particle velocities have 
$\prob(v) \propto \exp(-\sqrt{2} \mod{v}/ \sigma_v)$.  
Then, $\delta(\k)$ is multiplied by 
$(1 + k^2 \mu^2 \sigma_v^2 / 2)^{-1}$.

\item {\bf Pairwise exponential}.  In this case, the 
{\em pairwise} velocity dispersion of the particles is
assumed to come from an 
exponential distribution like that of the exponential model.  
The multiplicative factor is the square 
root of that in the previous case, and so the 
power spectrum itself ($\mod{\delta(\k)}^2$) is multiplied by 
$(1 + k^2 \mu^2 \sigma_v^2 / 2)^{-1}$.  
\end{enumerate}
The value of $\sigma_v$ that appears in the pairwise 
formula is the pairwise velocity dispersion, equal to 
$\sqrt{2}$ times the point-wise dispersion.  
Thus, if we Taylor expand each of the three factors, 
we find that the first order term is the same in 
each case; the distributions have the same width, 
but different shapes. 

\begin{figure}
\centering
\centerline{\epsfxsize= 8 truecm \epsfbox{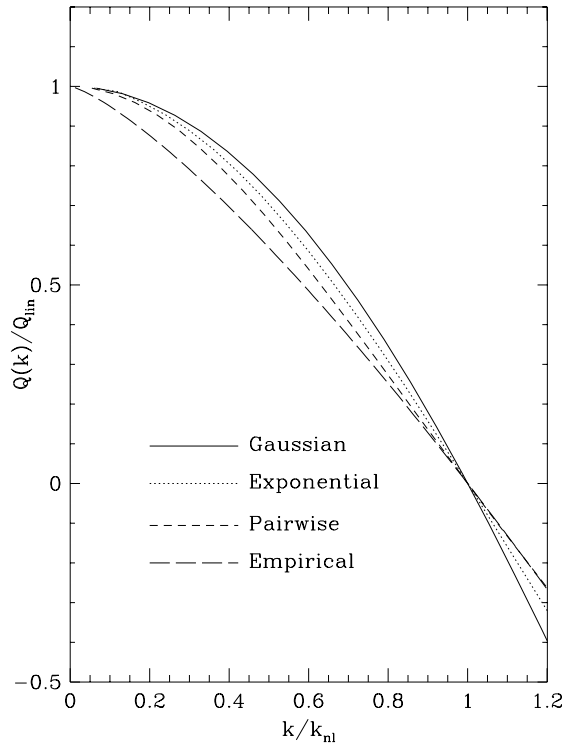}}
\caption{Comparison of our empirical model for $Q(k)$ with 
results from a Gaussian, exponential, and pairwise exponential 
model.  These models were computed for $\beta= 0.55$ and 
a one-dimensional point-wise 
velocity dispersion of $400\kms$.  }
\label{fig:compare_q}
\end{figure}

As pointed out by \citeN{PD94}, this effect cannot 
be considered in isolation; the linear Kaiser boost to the 
power spectrum also contains terms in $\mu$, and so to extract 
the harmonic moments, $P_0(k)$ and
$P_2(k)$, one must first multiply these two factors together before
weighting by the appropriate Legendre polynomial and averaging over
the line-of-sight angle.  We 
use the {\sc maple} computer package to perform these calculations.
In Fig.~\ref{fig:compare_q} we compare the three different 
velocity dispersion model predictions for the shape of 
$Q(k)$ with the shape we obtain for our 
empirical model, which we know (from Fig.~\ref{fig:scale}) is a 
reasonable fit to the simulations.  It is clear that the shapes 
of the velocity dispersion models do not match 
that of the empirical model, and that these models always 
underestimate the level of deviation from the linear 
theory result on large scales.  This breakdown is due to the 
violation of the assumption that non-linear behaviour arises purely 
through thermal motions in virialized systems at a fixed 
velocity dispersion.  In fact, the velocity dispersion will be 
correlated with the density field, and coherent non-linear flows may play 
an important part at large scales.

It is clear that using the velocity dispersion 
models to measure $\beta$ will result in some degree 
of bias, since the curves are not a good fit to the 
\nbody data.  In order to assess the significance of this bias  
we use the FFT method of estimating the redshift-space power spectrum
and its quadrupole-to-monopole ratio  
from the simulation cubes, as outlined above and described in HC98, for 
the full set of simulations described in CHWF.

\begin{figure}
\centering
\centerline{\epsfxsize= 8 truecm \epsfbox{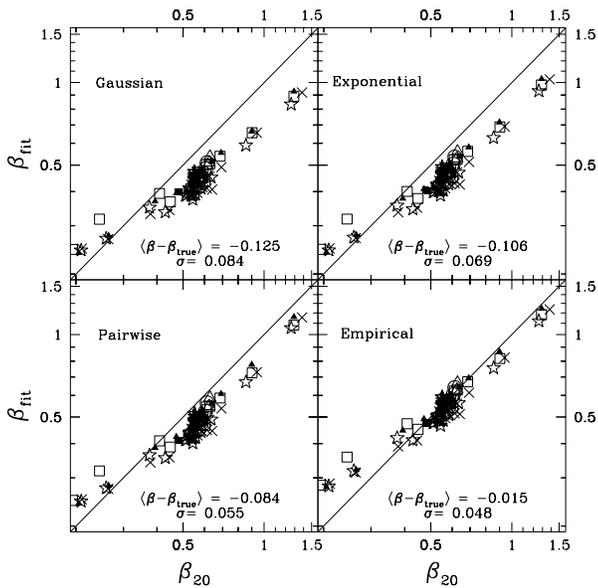}}
\caption{Scatter plots of $\beta_{\rm fit}$ for a sample of 
over one hundred simulations.  Each panel shows the results of 
using a different redshift-space distortion model to extract 
the value of $\beta_{\rm fit}$, as indicated by the name in the 
top left corner.  We also quantify the systematic offset, 
$\langle \beta - \beta_{\rm true} \rangle$, and the scatter, $\sigma$, 
about the mean value.  }
\label{fig:scat1}
\end{figure}

In Fig.~\ref{fig:scat1} we show a scatter plot for the 
behaviour of $\beta_{\rm fit}$ with  the known value of 
$\beta_{20}$, as defined above.  Each panel is labelled 
with the dispersion model used to make the fit, and the 
figure includes every biasing scheme of every cosmological 
model presented in CHWF.  
It is immediately clear from this figure that the velocity 
dispersion models all produce a systematic offset between
the fitted and true values of $\beta$.  The pairwise 
exponential model comes closest to fitting the data, 
but this still results in a systematic underestimate 
of $\beta$ by $\approx 0.1$.  Judging from a number of 
recent results (for a summary, see table~1 of 
\citeNP{Hamil97}), we expect to find $\beta \sim 0.5$,
so this represents a $20\%$ bias in measuring $\beta$, 
certainly rather larger than the random errors expected 
from large surveys like 2dF and SDSS.  The estimator based
on our empirical model, shown in the bottom right-hand panel, has
similar scatter to the other estimators but there is no
appreciable systematic offset over a wide range of $\beta$.  
The only regime where the model appears to break down is for 
$\beta \lsim 0.3$, which is at the very low end of the range 
allowed by current observations.

\subsection{Intrinsic scatter}
In addition to the original two $\tau$CDM simulations (E3SA and 
E3SB of CHWF), we now have a further eight simulations of the
same cosmological model. All ten of these realizations have 
been biased with one of the bias prescriptions.  The scatter in the
resultant $\beta_{\rm fit}$ values of $\sigma=0.03$ is an
indication of the limiting accuracy to which $\beta$
can be determined due to the finite volume of these
simulations.  This result indicates that the offsets in
$\langle \beta - \beta_{\rm true} \rangle$ seen for the
empirical model fits in Figs.~\ref{fig:scat2} and~\ref{fig:scat1} 
are statistically insignificant.

\section{Analysis of redshift surveys}
\label{sec:mock}
Having demonstrated that our empirical model for the
quadrupole $Q(k)$ can be used to make essentially
unbiased estimates of the redshift-space distortion parameter
$\beta$, we now turn to the question of how accurately
$\beta$ can be measured from a real galaxy redshift survey.
We choose to apply the technique to the geometry of the 2dF  
South Galactic slice \cite{Coll95}.  

First, in section~\ref{sec:nbody}, we address this question 
by making use of mock galaxy catalogues drawn from the ten independent
biased $\tau$CDM \nbody simulations mentioned above.
Then, in section~\ref{sec:ideal}, we perform a similar analysis
of idealized mock catalogues, in which the galaxy density 
field is assumed to be a Gaussian random field. 
The results are compared in section~\ref{sec:comp}.

\subsection{\nbody mock catalogues}
\label{sec:nbody}

The ten mock 2dF galaxy catalogues that we analyse here have
many realistic features.  They are based on \nbody simulations
in which the small scale structure has been accurately evolved into
the non-linear regime.  
The galaxies are biased tracers of the underlying dark matter, 
and have a correlation function that, on small scales, matches 
that of the galaxies in the APM survey.  The 2dF South Galactic 
slice is approximately $90\degr$ long in right ascension, with 
declination range $15\degr$ centred on $\delta=30\degr$.  
Our mock catalogues have the same geometry on the sky,   
and are constructed with a similar radial selection function to that 
expected from the real survey.  
The way in which we deal with the wide opening angle of the survey 
and select the galaxy sample to analyse is outlined below.

\subsubsection{Distant observer approximation}
\label{sec:slices}
The Cartesian linear-theory formalism of \citeN{Kaiser87} is 
applicable only when the line-of-sight direction is constant
across the whole survey.  In this case, when the redshift-space
density field is decomposed into the sum of plane waves, 
only the components of the waves parallel to the line of sight
are affected by the redshift-space distortion \cite{ZH96}.  
Thus it is not directly applicable when the galaxy survey 
subtends a large angle on the sky.  Whilst there 
do exist methods of dealing with wide angle surveys in 
one go, by expanding the density field into spherical harmonics 
\cite{HT95}, 
we do not concern ourselves with those here.  Rather, we 
split the survey up into separate angular bins with 
relatively small opening angle, and treat each one with 
the plane parallel approximation.  For the 2dF SGP geometry, 
the declination range is $15\degr$.  It thus seems 
reasonable to pick a right ascension range for the  
angular bins of comparable extent.  The width 
of the South Galactic strip is $\approx 90\degr  
\cos 30\degr = 78\degr$.  We split the survey into three 
bins each spanning $26\degr$ in right ascension.  Note that 
there are pairs of galaxies that span adjacent bins but still 
have opening angle less than the $26\degr$ bin width.  To 
avoid throwing this data away, we also include the two 
overlapping bins.  The effect of a finite 
opening angle on the estimated value of $\beta$ has been 
studied by \citeN{CFW95}.  From their fig.~8 it 
is clear that a $26\degr$ opening angle should only result 
in a one per cent underestimate of the quadrupole-to-monopole 
ratio of the power spectrum.  Since $Q_{\rm lin}(\beta)$ 
is, from equation~\ref{eqn:qlin}, approximately linear over a 
reasonable range of $\beta$ (estimates of $\beta$ range from 
around $0.2$ to $1.0$ \cite{Hamil97}), 
this introduces only a minor systematic error.  
We thus perform the analysis for five bins of angular extent   
$\approx 26\degr \times 15\degr$ and average the resulting 
power spectrum estimates.

\subsubsection{Volume limited samples}
The great advantage of surveys with such  depth and 
sampling as 2dF and SDSS is the ability to construct 
volume limited samples of galaxies.  We can thus look 
at the clustering properties of a particular class of 
galaxies, avoiding the problems that arise if
bias is a function of luminosity.  

In a volume limited sample, we throw away all the galaxies 
beyond a certain redshift limit, and all galaxies with 
absolute luminosities such that they would not make it into 
the catalogue if they were located at that redshift limit.  

For the genuine survey it will be interesting to carry out analysis 
of redshift-space distortions for a set of volume limited catalogues 
of varying depth. Here we simply illustrate the analysis for one 
volume limit chosen to be close to optimal for determining $\beta$.  
As the volume limit is increased we retain only the brighter 
galaxies, and so the galaxy number density decreases and the shot-noise 
contribution to the power spectrum increases.  For an accurate 
constraint on $\beta$, we should choose a depth for which the survey 
will give reliable data, without errors due to shot noise 
dominating over the contribution due to the finite size of the survey 
volume.  For the purposes of illustration, we have chosen this depth to be 
such that the resulting shot-noise contribution to the power spectrum 
is equal to that of the clustered galaxy distribution at $k = 0.3 \himpc$, 
so that on larger scales the error on $P(k)$ is not shot-noise dominated.  
The reason for this choice of scale is simply that we are going to fit 
our model for the quadrupole distortion only as far as the 
zero-crossing scale $\knl$, which, for the model under investigation, 
is $\knl \approx 0.3 \himpc$.  
The resulting redshift limit arising from this procedure is 
$z \approx 0.29$.  The median redshift of the survey is rather 
lower than this,  $\zbar \approx 0.1$.  

At this limit, the survey is very sparse, with $\nbar \approx 5 
\times 10^{-4} \hhhimpc$.  We stress that this sacrifice in 
density has been made in order to probe the longest, most linear 
scales as accurately as possible, and that the resultant high 
shot noise has little or no detrimental effect on measurements 
of the long wavelength, high amplitude modes.  

In catalogues with this depth, we expose ourselves to a number 
of factors which can contaminate the clustering signal.  The  
ten mock catalogues considered here are constructed from simulations with 
$\Omega_0=1,\Lambda_0=0$.  In general, the mapping from redshift 
to comoving distance is a function of the cosmology, and so,  
if the wrong cosmology is assumed in this transformation, 
the size of a volume element will appear to change with redshift, 
essentially mixing the measured clustering between scales.  This 
effect is only expected to be noticeable for high redshift samples, 
$z \gsim 0.1$, and so a sample of this depth would require 
investigation to determine the magnitude of potential bias 
given the uncertainty in cosmology.  
Inherent in having a significant look-back time is the possibility for 
significant evolutionary effects within the sample.  As the density 
field evolves under gravitational instability, the strength of  
clustering increases.  Conversely, if the galaxies are biased 
tracers of the density field, the bias factor approaches unity 
as evolution progresses \cite{Fry96}, so the resultant effect 
on galaxy clustering is quite non-trivial to compute.  Furthermore, 
we should allow for the possibility that the galaxies can themselves 
evolve, so a sample of galaxies between certain luminosity bounds 
at high redshift will not necessarily have the same clustering properties 
as a similar sample in the local Universe.  

The interplay between these factors is complex, and we make no attempt 
to deal with any of these effects here, assuming 
that they are either insignificant, or that they can be accurately 
corrected for in real datasets.  
In section~\ref{sec:mag} we will discuss the increase in accuracy 
that comes about if we deal with the whole, magnitude limited 
catalogue rather than a volume limited subsample.  

\subsubsection{Estimates of $\beta$}
\label{sec:nbody_e}
In order to estimate $\beta$ for each mock catalogue we first
estimate the power spectrum and $Q(k)$ for each catalogue using
a standard FFT method. To perform the fit we also need an estimate
of the error on each measurement of $Q(k)$ and ideally the covariance
of the estimates of $Q(k)$ at different values of $k$.  
These covariances arise through the finite size of the window 
function, and we estimate the full covariance matrix introduced by 
this effect using the method of multiple idealized mock catalogues 
that will be described  in the following section.  
We then use this matrix to make generalised minimum-$\chi^2$ 
fits of the empirical model of $Q(k)$ to each of the ten realizations.  
This process results in an estimate of $\beta$ for each of the mock 
catalogues.  

The true value of $\beta$ in these simulations is (the stars 
in Fig.~\ref{fig:scat2}) $\beta_{\rm true}=\beta_{20} = 0.55$,
while from the mock catalogues we find an average value of
$\beta_{\rm av} = 0.493$, with a scatter of $\sigma = 0.173$.  
First we note that the difference between $\beta_{\rm true}$
and $\beta_{\rm av}$ is not statistically significant given
that the standard error on the mean of the ten realizations is
$0.173/\sqrt{10} = 0.055$.  The scatter on $\beta$, however, is 
around $35\%$ of the mean; this is far in excess of the 
uncertainty we hope to achieve with a survey of the 
size of 2dF.

\subsection{Idealized linear mock catalogues}
\label{sec:ideal}

We wish to isolate the reason why the scatter in the estimated
$\beta$ values is substantially larger than expected. To do this
we have generated a large set of idealized mock catalogues in which
the galaxy density is explicitly assumed to be a Gaussian random
field rather than the result of biasing an evolved \nbody simulation.

We construct the density field on a cubic grid by first obtaining  
its Fourier transform, using a simple model for the redshift-space 
power spectrum $P^S(k,\mu)$ and random phases.  We choose one axis to 
be the line-of-sight direction and so assume that it is constant 
across the whole survey.  This means we avoid the problem of large 
opening-angle discussed in section~\ref{sec:slices}, and if we wish 
we can analyse the whole survey volume in one piece rather than having 
to slice it into five wedges of smaller opening angle.  The difference 
in accuracy between these two approaches indicates the scope 
for improving the estimates of $\beta$ by carrying out the analysis 
using spherical harmonics, which in principle allow the effects of a 
large opening angle to be treated precisely.

The redshift-space power spectrum we adopt can be expressed as
\[
P^s(k,\mu) = P_L(k) K(\mu) E(k,\mu) + \pshot.
\label{eqn:pkz}
\]
Here $P_L(k)$ is the isotropic real space power spectrum which
we assume to be given by the CDM power spectrum of
\citeN{BBKS}:
\[
P_L(k) \propto \frac{k^n \times [\ln(1+2.34q)/2.34q]^2}{[1+3.89q+(16.1q)^2+(5.46q)^3+(6.71q)^4]^{1/2}},
\]
where $q=k/\Gamma$.  As with the galaxy power spectrum of the 
\nbody catalogues we set $n=1$, $\Gamma=0.25$, as suggested
by observations of large-scale structure \cite{PD94} and fix the
amplitude such that  $\sigma_8=0.96$, the value determined for 
galaxies in the APM survey \cite{MES96}.  The second factor,
$K(\mu) = (1 + \beta \mu^2)^2$, is the linear theory redshift-space 
distortion of \citeN{Kaiser87}. The effect of small scale velocity dispersion
is modelled using the pairwise exponential dispersion model by setting
$E(k,\mu) = [1 + (k \sigma_v \mu)^2/2 ]^{-1}$ with $\sigma_v = 500 \kms$.  
This is the dispersion model that comes closest to fitting the
shape of $Q(k)$ in the \nbody catalogues and the value of
$\sigma_v$ adopted produces a zero-crossing of $Q(k)$ at close to the
$\knl \approx 0.3\himpc$ found in those catalogues.
The final term in equation~\ref{eqn:pkz} 
represents the shot noise in the catalogue
and we set $\pshot = 1/\nbar$, where $\nbar$ is the  galaxy
number density in the volume limited catalogues of  
section~\ref{sec:nbody}.  

This model ensures that the power spectrum of these 
idealized galaxy density fields matches well those of the
volume limited \nbody mock catalogues analysed above.   
We now give each mode a random complex phase and 
use an FFT to produce the galaxy density field in a cube.   
This field is then sampled using the survey window function.  
We can then either split this volume into five wedges or analyse 
the whole volume in one piece.  We can also vary the size of the 
cube in which the density field is constructed to assess how this 
effects the constraint on $\beta$.

\subsubsection{The survey window function}

The effect of sampling the galaxy density field only inside 
certain angular boundaries and with a particular radial selection 
function can be modelled by multiplying the full galaxy density 
field, $\delta(\r)$, by a spatial window function, $W(\r)$. The 
effect on the Fourier transform of the density field is that of a 
convolution with the transform $W(\k)$, and similarly the power 
spectrum is convolved with $|W(\k)|^2$.  This process can be 
thought of as a smoothing of the three-dimensional power spectrum, 
$P(\k)$, with an anisotropic window function $|W(\k)|^2$. On large 
scales (small $k$) the effect of this is to suppress the 
quadrupole-to-monopole ratio.  In fitting our models to the $Q(k)$ 
measured from the mock surveys we take account of this by first 
explicitly convolving the model  power spectrum with the survey 
window function.  We do this numerically using FFTs and the 
convolution theorem.

\subsubsection{Estimates of $\beta$}

As with the \nbody mock catalogues we estimate
$Q(k)$ for each realization of an ensemble of mock catalogues. 
Here we can use many more than ten and so determine the mean, variance
and covariance of the values of $Q(k)$ very accurately.
We use the resultant covariance matrix  to perform a fit to $Q(k)$
in each realization over the same range of $k$.
In this case the model for $Q(k)$ that we fit is that for the
analytic pairwise exponential model that we used to construct the 
power spectra. Thus we are again in the idealized realm of having a
completely accurate model for the redshift-space distortions, 
but we have shown that the empirical model is indeed a good fit to 
the \nbody simulation data.  Both models for $Q(k)$ have two 
parameters which control the large-scale asymptote and the 
zero-crossing, so it is reasonable to assume that confidence 
limits on $\beta$ derived using one model will be similar to 
those using the other.

The resulting scatter in $\beta$ when we analyse the idealized
mock catalogues split into the five wedges as we did for the
\nbody mock catalogues is $20\%$.  This result is almost a factor 
of two smaller than the scatter we found in section~\ref{sec:nbody_e}.

If we analyse the whole survey coherently, without splitting it up 
to satisfy the distant observer approximation, then we find a scatter 
of $16\%$.  This change from $20\%$ to $16\%$ represents the modest
improvement that can be made if one accurately deals with the
complications of the effects of the large opening angle of the
2dF survey.

The effect of increasing the simulation box size to the point 
where it is much larger than that used for the \nbody simulations
and much larger than the depth of the 2dF survey is to reduce the
scatter from $20\%$ to $18\%$.  Thus little information is
lost due to the finite volume of the \nbody simulations used to create
the mock catalogues.

\subsection{Discrepancy}
\label{sec:comp}

We now address the question of why the mock catalogues based on the
\nbody simulations produced an error on the estimated value of
$\beta$ almost twice as large as the idealized analysis predicts.

\begin{figure}
\centering
\centerline{\epsfxsize= 8 truecm \epsfbox{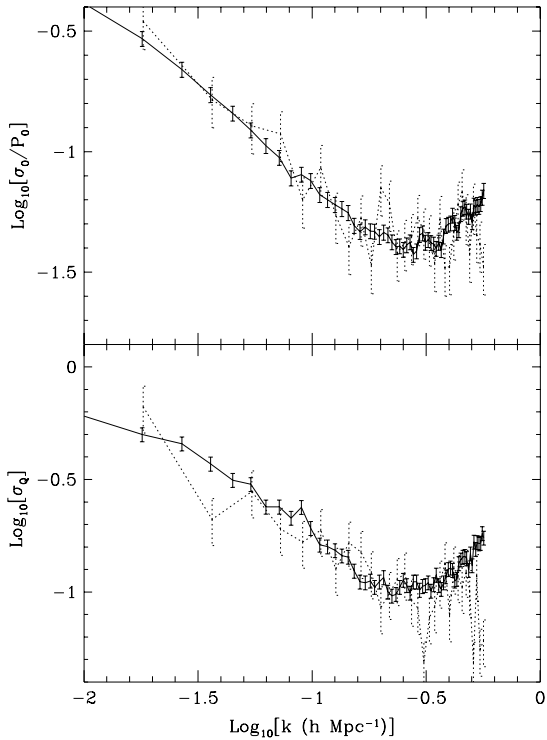}}
\caption{The fractional error on the power spectrum (upper panel)
and absolute error on the quadrupole-to-monopole ratio (lower panel).  
The solid line is from the variance between then ten independent 
\protect\nbody mock catalogue realizations of the survey, the dotted 
line is from one hundred `linear' mocks as described in 
section~\protect\ref{sec:ideal}.  We have attached 
Poisson error bars to the points using the relation that the standard 
deviation of the scatter is given by 
$\sigma_{\rm ss} = \sigma_{\rm s} / \sqrt{2(N-1)}$, 
where $\sigma_{\rm s}$ is the measured scatter from 
$N$ realizations \protect\cite{Barlow}.}
\label{fig:frac_errs}
\end{figure}

Fig.~\ref{fig:frac_errs} shows the fractional error on $P(k)$ and 
the absolute error on $Q(k)$ as a function of wavenumber.  It compares
the variance from the ten independent \nbody mock catalogues with 
the error we find from the idealized Gaussian random fields.  
The correspondence between the two error estimates clearly 
indicates that the Gaussian random field method has not under-predicted 
the degree of uncertainty on the distortion of each individual  
mode.  We also note that these error estimates are in good agreement 
with the values predicted by the method of error analysis detailed 
in \citeNP{FKP94}.  
We therefore conclude that the reason for the difference in the errors 
in $\beta$ must be due to the existence of 
significant correlations between $Q(k)$  at different values of $k$.  
These correlations are over and above those that are induced 
by  the shape of the survey window function, 
which we have taken into account in both methods.  
Therefore they must be a result of non-linear gravitational evolution 
and so cannot be modelled by a Gaussian density field.  

Gravitational instability on small scales causes coupling  
between different modes of the density field, resulting in 
correlations between the phases of these modes \cite{MW99}.    
We fail to take into account this non-linear effect when using 
idealized, linear mocks, but it is implicit in the \nbody method 
which has followed the evolution of these modes accurately.  
We only expect linear theory to be valid for $\Delta^2(k) \ll 1$, 
where $\Delta^2(k) = 4\pi/(2\pi)^3 k^3 P(k)$, and this condition is 
certainly not met in our \nbody simulations around the zero-crossing 
of the quadrupole.   However, the strength of 
the effect is perhaps initially surprising.  

As a first attempt to deal with these non-linearities, we 
repeat the \nbody fits described in section~\ref{sec:nbody_e}, but replace 
the covariance matrix used previously with a new matrix that is 
obtained from the ten realizations themselves.  The off-diagonal 
terms in this matrix arise from mode coupling caused by the survey 
window function and also non-linear evolution of the density field.  
The error in the estimate is now reduced from $35\%$ 
to $25\%$.  This reduction occurs because previously we attached 
much weight to the quasi-linear regime, to modes at  intermediate wavelengths 
close to, but longer than, $\knl$, where the quality of data appeared
good.  In actual fact, this regime is also the scene of strong non-linear 
mode coupling, so the data are not in fact as good as previously 
believed.  Applying the more realistic weighting scheme 
makes better use of the data, and hence produces less scatter in the 
estimate of $\beta$.  
Unfortunately, this method of obtaining the covariance matrix is 
time-consuming, since it involves producing multiple 
large-volume \nbody simulations 
for each cosmology and power spectrum that is to be studied.   

\subsection{Summary}
In sections~2 and~3, we have:
\begin{itemize}
\item used biased \nbody {\it simulations} to develop an accurate, 
empirical model for $Q(k)$.
\item used {\it mock catalogues} drawn from these simulations 
to illustrate how accurately $\beta$ can be measured in practice.  
\item used {\it linear} mock catalogues to show how much of the 
uncertainty in $\beta$ comes from just the finite volume effect.  
\item demonstrated that, in fact, a major source of random error 
is the coupling of non-linear modes.  
\item shown that allowing for this coupling (by using a covariance 
matrix derived from the {\it full} mock catalogues, rather than the 
{\it linear} mocks) results in a large improvement in constraining 
$\beta$.

\end{itemize}

The error has been reduced by allowing for non-linear effects, 
but is still rather large, in excess of the level of 
constraint that could be achieved if non-linearities were not present 
at all.  In the next section, we outline a practical method for reducing the 
non-linearities and thus increasing the accuracy of the estimates.  
It is also worth mentioning the approach adopted by \citeN{Hamilton99}, 
who finds a method for constructing a near minimum variance estimator 
for the non-linear power spectrum based on assembling a set of 
almost uncorrelated non-linear modes.  The use of such optimal 
methods in both the non-linear and linear regimes may provide the 
capability to constrain $\beta$ rather better than we do in this work, 
although the methods are still in the development stage and it is not 
clear how easily they can be applied to a real dataset.

\section{Collapsing clusters}
\label{sec:clus}
In the dispersion models one assumes that the random velocities 
induced by non-linearities are uniform and uncorrelated with the 
density field. In reality we do not expect this to be the case; 
the coupling between $k$-modes is a local effect in $r$-space, 
taking place inside particular non-linear structures.  In these 
volumes, mode coupling is extremely strong.  As pointed out in 
HC98, if we can somehow excise these regions from our treatment, 
we could create a better behaved, more linear quadrupole-to-monopole 
ratio.  In addition, this would have an even more dramatic effect in 
lessening the correlations between modes on these scales.  This 
result could be achieved by identifying and removing the signal from 
clusters, as described in HC98.

Here we illustrate the application of this method.  Taking the 
ten simulations used earlier, we identify clusters in real space using a 
friends-of-friends \cite{DEFW} algorithm with linking length $0.2$ 
times the mean inter-particle separation.  We select clusters with 
ten or more members, and compute the mean velocity of each 
cluster by averaging over the velocities of its members.  
The cluster members then have their velocities in the 
simulation replaced with that of their parent cluster.  
In this way, it was shown in HC98, the effect of non-linear 
velocity dispersions can be heavily damped.


In a real survey, clusters will have to be measured 
without the perfect knowledge of galaxy density field 
which we have assumed here, and in redshift space 
rather than real space.  This is likely to involve 
a number of parameters and considerations, and from 
this point of view it seems pointless to develop 
an accurate model for the cluster-collapsed $Q(k)$ since 
this will not be generally applicable.  Here we instead take as 
a model for $Q(k)$ the shape measured from averaging  
the ten full-cube simulations.  We multiply this by a 
single free parameter, $Q_{\rm lin}$, which 
we assume is related to $\beta$ via the linear theory 
result (equation~\ref{eqn:qlin}).  For the set of $\tau$CDM 
simulations, non-linearities are suppressed to the extent that the 
quadrupole now crosses the $k$-axis at $\knl = 0.8 \himpc$.   

Having derived this model from the simulations, we now apply it 
to mock catalogues.  These are created from the cluster-collapsed 
simulations, with the same parameters and selection functions 
as the mock 2dF SGP catalogues used in section~\ref{sec:mock}.  
We then apply the same FFT technique explained above 
to measure $Q(k)$ in the mock catalogues.  

We perform the fits over the same range of $k$ as before 
($0 < k < 0.3 \himpc$), using a covariance matrix obtained from 
the ten independent cluster-collapsed mock catalogues.  
We find that the scatter in $\beta$ is reduced from $25\%$
to $13\%$.  This is better than the result from the Gaussian random
fields and indicates that collapsing clusters has successfully 
removed most of the mode correlations on these scales.
Collapsing the clusters has pushed the zero-crossing 
of $Q(k)$ to smaller scales, so we may be able to reduce the uncertainty 
still further by extending the range over which we fit the model.  
However, this is not the case; extending the fits to 
$k = 0.6 \himpc$ results in no significant increase in accuracy.  
This is because the data in this range is much noisier than in 
the uncollapsed case, aiding little in fitting the the curve.  

We note that now we are using scales that were previously quasi-linear 
to determine $\beta$, so the fits will not be greatly improved by relaxing 
the small angle constraint.  Much of the new data is 
coming from pairs whose separation on the sky is small compared to
the full extent of the survey.

\section{A magnitude limited sample}
\label{sec:mag}
Throughout the above work we have used the same volume limited sample 
with $z_{\rm lim} = 0.29$ to measure $\beta$.  Volume limiting 
selects a subsample of galaxies, which can be very useful for studying 
clustering behaviour as a function of luminosity, but at the expense 
of discarding the majority of galaxies and hence severely reducing the 
number density of objects.  If we are interested in using the survey 
to measure $\beta$ as well as possible for all the galaxies in the 
catalogue, we can extend the treatment to the magnitude limited case.  
For the case of a real survey, this must be done with caution, 
since we know that galaxies of different luminosity exhibit different 
clustering properties (\eg \citeNP{LTM99}).  
In our mock catalogues, however,  there is no dependence of clustering strength 
on galaxy luminosity, so any sample should measure the same power 
spectrum and hence the same $\beta$.  A magnitude limited sample will 
be more accurate since the number of galaxies is higher, and it includes 
galaxies at higher redshifts, probing more long-wavelength modes.  
We weight galaxies using the scheme introduced by \citeN{FKP94}: 
\[
w(\r) = \frac{1}{1+\nbar(\r)P_{\rm w}}
\]
where $\nbar(\r)$ is the number density at position $\r$, and 
we set $P_{\rm w} = 5000\hhhimpc$, roughly reflecting the value of the 
galaxy power spectrum around the middle of the range over which the 
fit is performed ($k \approx 0.15 \himpc$).   This weighting is 
designed to minimize the variance in estimates of $P(k)$  where 
$P(k) \sim P_{\rm w}$.  

Using this method, applied to the cluster-collapsed catalogues described 
in the previous section, results in a reduction of the error on $\beta$ 
from $13\%$ to $10\%$.  A similar improvement is expected for the 
uncollapsed case.  

\section{Conclusions}
\label{sec:conc}
There are four distinct results presented in this paper:

\begin{enumerate}

\item We have found an empirical model for the quadrupole
moment, $Q(k)$, of the redshift-space power spectrum which
matches the behaviour of this quantity in a wide range 
of cosmological models much more accurately than existing
analytic models. We have demonstrated that it can be successfully
used to obtain unbiased estimates of the redshift-space
distortion parameter $\beta=\Omega^{0.6}/b$.

\item We have shown that the resultant measurement of $\beta$ 
is better correlated with the large-scale bias factor than the 
usual quasi-linear measure from spheres of radius $8\hmpc$.  
The quadrupole is thus not sensitive to small-scale 
scale-dependence of the bias.  

\item We have demonstrated that mode coupling caused by non-linear 
gravitational 
evolution causes strong correlations between the measured
values of  $Q(k)$ at different values of $k$, even on quite large scales.
This implies that estimates of the accuracy to which $\beta$ can be 
constrained that are based on assuming the galaxy density field is Gaussian
significantly under estimate this error.

\item The most non-linear regions of the density field are galaxy clusters.
Although these are compact structures in real-space they are very
extended in redshift space and so can affect quite large
scale modes in the redshift-space density field.
We have shown that by identifying and collapsing galaxy clusters
we are broadly successful at both reducing the correlation and increasing the 
range of scales over which $Q(k)$ can be reliably fitted.

\end{enumerate}

For a set of 2dF southern slice mock catalogues which have a true 
value of $\beta=0.55$, 
our empirical model of redshift space distortion produces unbiased 
estimates of $\beta$ with  a $25\%$ statistical error when measured 
from a volume limited sample with limiting redshift $z_{\rm lim} = 0.29$.  
This model is applicable to the raw, redshift-space galaxy 
distribution and does not require the identification or manipulation
of galaxy clusters.  We estimate the statistical error will be 
significantly reduced (to around $20\%$) if the survey is analysed in 
one piece, rather than being chopped into five overlapping 
wedges, which we do in order to satisfy a small angle constraint.

If clusters can be successfully identified and collapsed in
the 2dF galaxy redshift survey, analysis of the main southern slice
should constrain $\beta$ to a greater accuracy of $13\%$ (volume limited) 
and $10\%$ (magnitude limited).  This error is not expected to be 
significantly reduced by relaxing the constraint of the distant 
observer approximation and using galaxy pairs at large angular 
separation, since much more of the information now comes from 
pairs that are physically nearer each other, and hence closer 
together on the sky.  

It is worthwhile noting that \nbody simulations, when normalized 
to replicate the observed abundance of galaxy clusters, produce 
pairwise dark-matter velocity dispersions which are significantly 
higher than those measured for galaxies \cite{Jenk98}.  This problem 
is largely alleviated in new models which attempt to realistically
identify the places where galaxies form, either by numerical \nbody 
hydrodynamic simulation \cite{Jenk99}, or by semi-analytic modelling 
\cite{Benson99b}.  However, the simple bias models used in this
paper produce pairwise galaxy velocity dispersions which are close 
to those of the dark matter and therefore somewhat higher than for
observed galaxies.  It is to be hoped that this fact alone will 
render observational datasets somewhat less prone to non-linear 
effects than we estimate from our simulations, and the $10\%$ 
statistical error we find may be a little conservative.

The 2dF survey also contains a slice in the northern Galactic 
hemisphere.  This is a little smaller than the southern slice, but 
measuring $\beta$ from this data and combining it with the southern 
slice provides an additional constraint.  Assuming the errors add 
in quadrature, the uncertainty is reduced by a factor $\approx 1.25$.   

We thus conclude that the quadrupole moment of the redshift-space 
distortion measured in the 2dF survey will be capable of constraining 
$\beta$  to an accuracy of $8\%$.  This will yield an important 
constraint on the density parameter $\Omega_0$.

\section*{Acknowledgments}
SJH acknowledges the support of a PPARC studentship and funding 
from Durham University.  SMC acknowledges the support of a PPARC 
Advanced Fellowship.  The authors wish to thank the referee, 
Michael Strauss, for an excellent job in suggesting improvements 
to the clarity of this work.

\bibliographystyle{mnras}


\end{document}